\documentclass[]{aa}
\usepackage{graphicx,hyperref,txfonts}
\usepackage{caption}
\usepackage{subcaption}
\usepackage{color}

\newcommand{\com}[1]{\textnormal{#1}}
\newcommand{\comm}[1]{\textnormal{#1}}

\begin{document}

\title{Generalised model-independent characterisation of strong gravitational lenses III: perturbed axisymmetric lenses}
\titlerunning{Model-independent characterisation of perturbed Einstein-rings}
\author{J. Wagner\inst{1,2}}
\institute{Universit\"at Heidelberg, Zentrum f\"ur Astronomie, Institut f\"ur Theoretische Astrophysik, Philosophenweg 12, 69120 Heidelberg
\and
Heidelberg Institute for Theoretical Studies, 69118 Heidelberg, Germany \\
\email{j.wagner@uni-heidelberg.de}}
\date{Received XX; accepted XX}

\abstract{In galaxy-galaxy strong gravitational lensing, Einstein rings are generated when the lensing galaxy has an axisymmetric lensing potential and the source galaxy is aligned with its symmetry centre along the line of sight. Using a Taylor expansion around the Einstein radius and eliminating the unknown source, I derive a set of analytic equations that determine differences of the deflection angle of the perturber weighted by the convergence of the axisymmetric lens and ratios of the convergences at the positions of the arcs from the measurable thickness of the arcs. In the same manner, asymmetries in the brightness distributions along an arc determine differences in the deflection angle of the perturber \comm{if the source has a symmetric brightness profile and is oriented parallel to or orthogonal to the caustic}. These equations are the only model-independent information retrievable from observations to leading order in the Taylor expansion. General constraints on the derivatives of the perturbing lens are derived such that the perturbation does not change the number of critical curves. To infer physical properties such as  the mass of the perturber or its position, models need to be inserted. The same conclusions about the scale of detectable masses (of  the order of $10^8 M_\odot$) and model-dependent degeneracies as in other approaches are then found and supported by analysing B1938+666 as an example. \com{Yet, the model-independent equations show that there is a fundamental degeneracy between the main lens and the perturber that can only be broken if their relative position is known. This explains the degeneracies between lens models already found in simulations from a more general viewpoint. Hence, apart from the radii and brightness distributions of the arcs, independent information on the axisymmetric lens or the perturber has to be employed to disentangle the axisymmetric lens and the perturber. Depending on the properties of the pertuber, this degeneracy can be broken by characterising the surrounding of the lens or by measuring the time delay between quasar images embedded in the perturbed Einstein ring of the host galaxy. }
}

\keywords{cosmology: dark matter -- gravitational lensing: strong -- methods: data analysis -- methods: analytical -- galaxies: general -- galaxies: mass function}

\maketitle

\section{Introduction and motivation}
\label{sec:introduction}

\begin{figure*}[ht]
\centering
\begin{subfigure}{0.24\textwidth}
  \centering
 \includegraphics[width=0.7\linewidth]{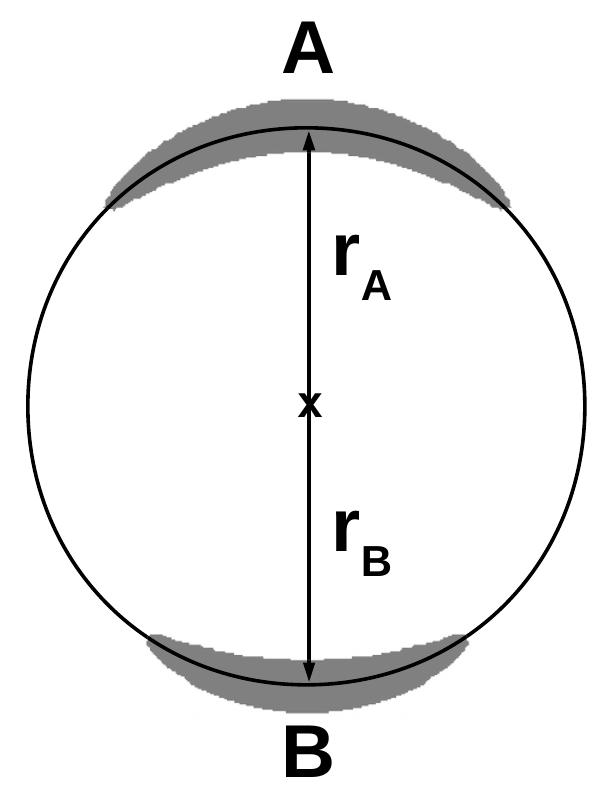}
 \caption{axisymmetric configuration}
 \end{subfigure}
 \begin{subfigure}{0.24\textwidth}
  \centering
 \includegraphics[width=0.7\linewidth]{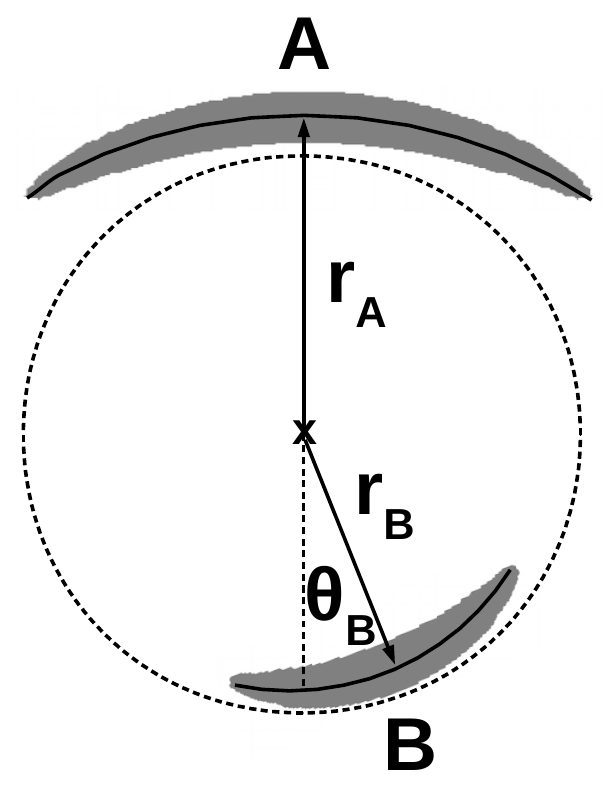}
 \caption{global perturbation}
 \end{subfigure}
 \begin{subfigure}{0.24\textwidth}
  \centering
 \includegraphics[width=0.7\linewidth]{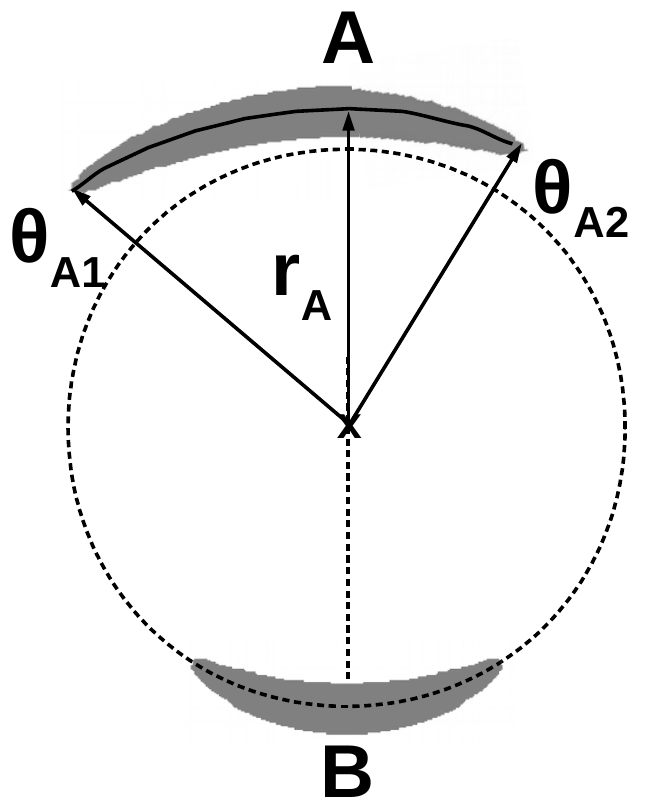}
 \caption{local angular perturbation}
 \end{subfigure}
 \begin{subfigure}{0.24\textwidth}
  \centering
 \includegraphics[width=0.7\linewidth]{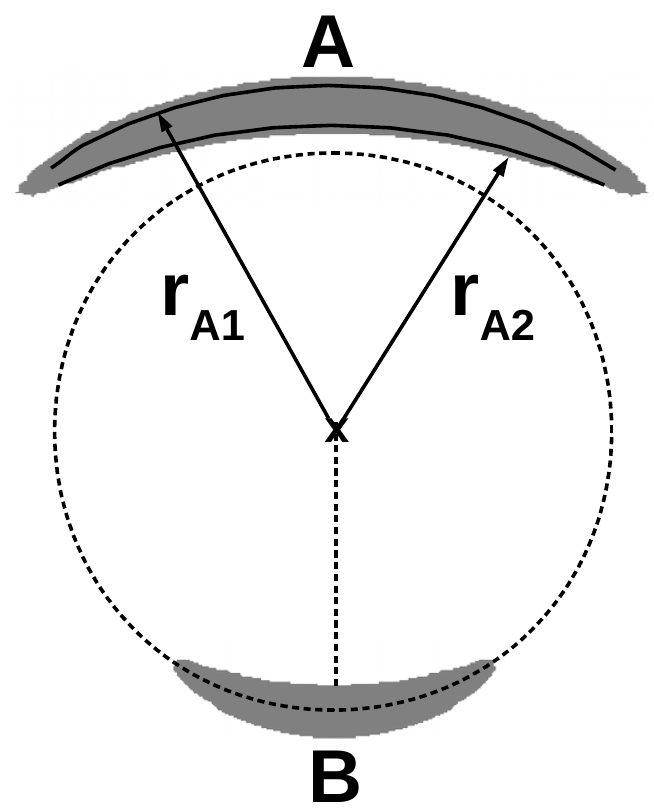}
 \caption{local radial perturbation}
 \end{subfigure}
    \caption{Possible configurations of the two multiple images $A$ and $B$ generated from a source lying close to an almost point-like caustic of an axisymmetric gravitational lens: without perturbations, $A$ and $B$ are aligned with the centre of the lens, and have symmetric arc lengths and a radius close to the Einstein radius (a); adding a global perturbation, the alignment symmetry between $A$ and $B$ can be broken (b); small, local perturbations can cause asymmetric arcs in the angular direction (c); or in the radial direction (d).}
    \label{fig:configurations}
\end{figure*}

In this third paper in the series  by \cite{bib:Wagner1} and \cite{bib:Wagner2}, I investigate which model-independent properties about a gravitational lens or the background source can be determined from two-image configurations caused by lensing mass distributions with perturbed axisymmetry. \cite{bib:Wagner1} determines local lens properties using the positions, shape, and orientation of multiple unresolved images close to a fold or cusp singular point on the critical curve of an arbitrary lensing potential. Subsequently, \cite{bib:Wagner2} extends the approach of \cite{bib:Wagner1} to multiple images with resolved structure in their brightness distribution. Given a resolved brightness structure in at least three multiple images of the same source, a transformation between these images can be set up and local lens properties at all image positions extracted.

A case with a topologically different image configuration that has not been analysed yet, is a projected, two-dimensional axisymmetric mass density with a small perturbation. The purely axisymmetric case is investigated in \cite{bib:Gorenstein}. \comm{In this work, I focus on} perturbed Einstein rings \comm{that} occur in \newpage \noindent galaxy-galaxy lensing and \comm{that} map an extended background object into two unresolved images on opposite sides of the lens centre. Possible perturbations are external shear or small-scale masses like black holes or satellite galaxies.

Examples of image configurations of perturbed Einstein rings, like B1938+666, can be found in the Sloan Lens ACS Survey (SLACS; \citealt{bib:Bolton}) or in the Strong lensing at High Angluar Resolution Program (SHARP; \citealt{bib:Lagattuta}). While the number of perturbed Einstein rings may still be small, they yield tighter constraints on the lensing mass distribution than less symmetric image configurations \cite[see e.g.][]{bib:Kochanek}. Hence, this lensing configuration is employed to investigate the missing satellite problem, to determine whether mass follows light, to calculate the values of cosmological parameters, or even to test alternative theories of gravity \cite[e.g.][]{bib:Alard, bib:Bolton, bib:Collett, bib:Suyu, bib:Vegetti2014, bib:Wertz}.

Existing  approaches for some of \comm{these applications were developed} in \cite{bib:Habara}, \cite{bib:Koopmans}, \cite{bib:Minor}, \cite{bib:Suyu}, and \cite{bib:Vegetti2009}. \comm{They employ parametric models or a combination of \newpage \noindent parametric and non-parametric approaches to describe the lensing configuration. Except for implementation details, all of them split the lensing configuration into a main lens and one or more perturbers, such as satellite galaxy or external shear due to the surrounding of the main lens. As a result, degeneracies arise as detailed in \cite{bib:Minor}, \cite{bib:Sluse}, and \cite{bib:Vegetti2014}. Within the framework of the model-independent approach used to characterise local properties of the gravitational lens, these degeneracies can be further investigated and understood from an analytical point of view. Subsequently coupling the model-independent approach with models, I show by means of an example, B1938+666, that the degenerate results are consistent with those obtained by the model-dependent approaches, for example the most probable configuration  found in \cite{bib:Vegetti2012}.}


In Section~\ref{sec:theory}, I describe the formalism and derive the constraints on the perturbing lensing potential that are inferred from the observables of the arc segments close to the critical curve. Subsequently, in Section~\ref{sec:limits}, I discuss the range of validity of the ansatz of a perturbed axisymmetric lensing potential and derive requirements for the derivatives of the lensing potential that need to be imposed on a perturber such that it does not alter the topology of the critical curves. \com{By deriving these limits, the multiple image configurations in perturbed Einstein rings are shown to be the limiting cases considered in \cite{bib:Wagner1} for merging images close to the cusp.} An analysis of \comm{the} occurring degeneracies is performed in Section~\ref{sec:degeneracies}. Then, I insert models for the axisymmetric and perturbing potentials in Section~\ref{sec:models}. Using these models, estimates on the scale of detectable perturbers are shown in Section~\ref{sec:substructure} and compared to the results of existing approaches. Section~\ref{sec:example} then shows B1938+666 as an example application before I summarise the results and conclude in Section~\ref{sec:conclusion}.


\section{Theory}
\label{sec:theory}

\subsection{Taylor-expanded lens mapping}
\label{eq:lens_mapping}

Let $\psi(\boldsymbol{x})$ be the deflection potential of a gravitational lens that defines the lens mapping between $\boldsymbol{x} \in \mathbb{R}^2$ in the image plane and $\boldsymbol{y} \in \mathbb{R}^2$ in the source plane by $\boldsymbol{y} = \boldsymbol{x} -\nabla \psi(\boldsymbol{x})$. In the following, I  use polar coordinates, i.e. $\boldsymbol{x} = (r, \theta)$ and $\boldsymbol{y} = (r_y, \theta_y)$. Assuming that the general deflection potential $\psi(r,\theta)$ is mainly governed by an axisymmetric part $\psi_\mathrm{a}(r)$ depending only on the radius $r$ from its centre, a source aligned with the lens centre along the line of sight will generate an Einstein ring as image at radius $r_\mathrm{E}$. 

Observable deviations from this Einstein ring are caused by a perturbing lensing potential $\psi_\mathrm{p}(r,\theta)$ with $\left| \psi_\mathrm{p}(r,\theta) \right| \ll \left| \psi_\mathrm{a}(r) \right|$. 
\com{The perturbing potential can represent a small mass in the vicinity of the main lens or an external shear component caused by the environment the lens is embedded in. In \cite{bib:Kovner} and \cite{bib:Kassiola}, for small ellipticities equivalence relations between elliptical mass distributions, elliptical potentials, and an axisymmetric model with external shear are established. Hence, given these equivalences, $\psi_\mathrm{p}$ can describe effective elliptical perturbations to the axisymmetric lens. While model-based approaches usually treat elliptical masses/potentials and external shear individually, the model-independent approach absorbs both in $\psi_\mathrm{p}$;  observations to characterise the environment of the lens are often not performed, so that the ellipticity of the mass/potential and the external shear remain degenerate.}

Without loss of generality, the origin of the coordinate system is set at the centre of the axisymmetric part of the lens.

Taylor expanding $\psi(r,\theta) = \psi_\mathrm{a}(r) + \psi_\mathrm{p}(r, \theta)$ around $r_\mathrm{E}$,
\begin{equation}
\psi(r, \theta) = \sum \limits_{n=0}^{\infty} \left( a_n + p_n (\theta) \right) (r - r_\mathrm{E})^n
\label{eq:Taylor_potential}
\end{equation}
with the coefficients $a_n$ and $p_n(\theta)$ determined as derivatives of the potential parts at $r = r_\mathrm{E}$
\begin{align}
a_n = \left. \dfrac{1}{n!} \dfrac{\mathrm{d}^n \psi_\mathrm{a}(r)}{\mathrm{d}r^n}  \right|_{r = r_\mathrm{E}}\;, \quad p_n (\theta) =  \left. \dfrac{1}{n!}  \dfrac{\partial^n \psi_\mathrm{p}(r,\theta) }{\partial r^n} \right|_{r = r_\mathrm{E}} \;,
\label{eq:coefficients}
\end{align}
the lens mapping reads
\begin{align}
r_y &= r - a_1 - p_1(\theta) - 2 \left( a_2 + p_2(\theta) \right) (r-r_\mathrm{E}) \label{eq:le1} \;, \\
\theta_y &= \theta - \dfrac{1}{r\color{black}{^2}} \left( \dfrac{\partial p_0(\theta)}{\partial \theta} +\dfrac{\partial p_1(\theta)}{\partial \theta} (r - r_\mathrm{E}) \right) \;, \label{eq:le2}
\end{align}
when taking into account terms at most linear in $r-r_\mathrm{E}$. A detailed derivation can be found in Appendix~\ref{app:derivations_les}.

If there is no perturber, i.e.\ $\psi_\mathrm{p} \equiv 0$, a source galaxy placed in the vicinity of $r_y = 0$ will be mapped to two giant arcs, denoted as $A$ and $B$, on opposite sides of the lens centre in the image plane. Without loss of generality, I assume arc $A$ to be of positive parity (outside the critical curve) and arc $B$ of negative parity (inside the critical curve). Angular coordinates $\theta$ are non-negative values measured from the centre of light of each arc to the desired position (see Figure~\ref{fig:configurations} (b) for an example), if not mentioned otherwise.

\subsection{Observables}
\label{sec:observables}

I use a segmentation along the isocontours of the intensity profile of the arcs to define their shape. Theoretically, the following observables per arc can be measured: 
\begin{itemize}
\item the position of its centre of light, 
\item its curvature, called the arc radius, 
\item the two radial segments from the centre of light to each end of an arc. They are also called half-lengths, as they are equal for \comm{sources with circularly symmetric or elliptical brightness profiles aligned or orthogonal to the caustic when they are} mapped by axisymmetric lenses. \end{itemize}

Depending on the extensions of the two-image configuration on the sky and the resolution of the observation, the shape of the two arcs is measured to varying accuracy and precision. \com{Details related to practical measurements, their accuracy, and precision are not taken into account until Section~\ref{sec:substructure}.}

Figure~\ref{fig:configurations} (a) shows the two-image configuration for an axisymmetric lens. To show that both images have the same curvature radius to good approximation, $r_A \approx r_B \approx r_\mathrm{E}$, the critical curve with radius $r_\mathrm{E}$ is shifted to lie on both images. Furthermore, it can be observed that the lens centre and the centres of the arcs are aligned, and that the arcs have symmetric half-lengths.

\subsection{Leading order constraints on the perturber}
\label{sec:equations}

Identifying two points $i$ and $j$ in the image plane with a common source coordinate, this source coordinate can be eliminated by subtracting the respective lens equations from each other. Hence, given two differing radii $r_i, r_j$ or two differing angles $\theta_i, \theta_j$, subtracting their Equations~\eqref{eq:le1} and \eqref{eq:le2} yields to leading order\begin{align}
r_i - r_j &= \dfrac{\alpha_{\mathrm{p}, r}(r_\mathrm{E}, \theta_i) -  \alpha_{\mathrm{p}, r}(r_\mathrm{E}, \theta_j)}{2(1-\kappa_\mathrm{a}(r_\mathrm{E}))} \;,\label{eq:ce1} \\
\theta_i - \theta_j &= \dfrac{\alpha_{\mathrm{p}, \theta}(r_\mathrm{E}, \theta_i)}{r_i\color{black}{^2}} -  \dfrac{\alpha_{\mathrm{p}, \theta}(r_\mathrm{E}, \theta_j)}{r_j\color{black}{^2}} \;, \label{eq:ce2}
\end{align}
using the convergence of the axisymmetric part of the potential at $r_\mathrm{E}$, $\kappa_a(r_\mathrm{E})$, and the components of the deflection angle $\alpha_\mathrm{p}$ caused by the perturber $\psi_{\mathrm{p}}$ in radial and angular direction.
A derivation of the expressions can be found in Appendix~\ref{app:derivations_ces}. 

Hence, as expected, Equation~\eqref{eq:ce1} states that any difference in the radii is caused by a difference of perturbing deflections in radial direction. Analogously, Equation~\eqref{eq:ce2} states that any difference in the angular positions is linked to a difference of perturbing deflection in angular direction. 

These two general equations describe two possible perturbations, which will be discussed in the following and are summarised graphically in Fig.~\ref{fig:configurations}:
\begin{enumerate}
\item[(b)] a global perturbation causing an asymmetry between the two arcs;
\item[(c)] a local perturbation causing an angular asymmetry between the delineating isocontours of one arc.
\end{enumerate}


\subsubsection{Global displacement perturbation}
\label{sec:case_b}

In this case, I specify $r_i$ and $r_j$ to be the radii determined at the centres of light of the arcs $A$ and $B$, respectively. Equations~\eqref{eq:ce1} and \eqref{eq:ce2} then read
\begin{align}
r_A - r_B &= \dfrac{\alpha_{\mathrm{p}, r}(r_\mathrm{E}, \theta_A) -  \alpha_{\mathrm{p}, r}(r_\mathrm{E}, \theta_B)}{2(1-\kappa_\mathrm{a}(r_\mathrm{E}))} \;, \label{eq:perturber_b1} \\
- \theta_B &= \dfrac{\alpha_{\mathrm{p}, \theta}(r_\mathrm{E}, \theta_A)}{r_A\color{black}{^2}} -  \dfrac{\alpha_{\mathrm{p}, \theta}(r_\mathrm{E}, \theta_B)}{r_B\color{black}{^2}} \;.
\label{eq:perturber_b2}
\end{align}
As a consistency test, I insert $r_A = r_B$ and $\theta_B = 0$, i.e.\ the centre of $B$ is aligned with that of $A$ as shown in Figure~\ref{fig:configurations} (a). The equations yield an axisymmetric perturber, thus no perturber to an axisymmetric lens, in accordance with the expectations.

\subsubsection{Local angular perturbation}
\label{sec:case_c}

Assume \comm{a symmetric brightness profile of the source with its symmetry axes aligned or orthogonal to the caustic, so that symmetric multiple images appear. For instance, an unstructured early-type elliptical galaxy that can be identified by morphology and kinematic properties can fulfil these criteria. Then, if} two different half-lengths at each side of the centre of an arc are observed, i.e. $r_A \theta_{A1} \ne r_A \theta_{A2}$, as indicated in Figure~\ref{fig:configurations} (c), I insert $r_i = r_j = r_A$ and $\theta_i = \theta_{A1}$, $\theta_j = \theta_{A2}$ into Equations~\eqref{eq:ce1} and \eqref{eq:ce2} to obtain
\begin{align}
\alpha_{\mathrm{p}, r}(r_\mathrm{E}, \theta_{A1}) &=  \alpha_{\mathrm{p}, r}(r_\mathrm{E}, \theta_{A2}) \label{eq:perturber_c1} \;, \\
r_A\color{black}{^2} \color{black}\theta_{A1} - r_A\color{black}{^2} \color{black}\theta_{A2} &= \alpha_{\mathrm{p}, \theta}(r_\mathrm{E}, \theta_{A1}) - \alpha_{\mathrm{p}, \theta}(r_\mathrm{E}, \theta_{A2}) \;. \label{eq:perturber_c2}
\end{align}
Assuming $\theta_{A1} = \theta_{A2}$, the axisymmetric case is retrieved. 
\comm{To ascribe differing half-lengths in one arc to a local angular perturbation, the brightness profile of the source has to be symmetric and oriented parallel or orthogonally to the caustic, which can be corroborated if the other arc has a symmetric shape. For sources with unknown morphology and orientation, a comparison between the two images $A$ and $B$ is therefore necessary before using Equation~\eqref{eq:perturber_c2}.}

\comm{As this perturbation only employs differences between observables in one arc, it could theoretically be investigated for single arcs as well. In this case, however,  the only conclusion to be drawn is that there is no reason to assume a perturbing potential when a symmetric arc is observed.
}

\subsubsection{Local radial perturbation}
\label{sec:case_d}

Next, I assume that the arcs are broad enough to observe two different radii on their centre-far and centre-close intensity isocontour, measured at the angular position of the centre of the arc. Figure~\ref{fig:configurations} (d) depicts this configuration for arc $A$, denoting the centre-far and centre-close radii measured at $\theta_A$ as $r_{A1}$ and $r_{A2}$, respectively. Then, Equations~\eqref{eq:ce1} and \eqref{eq:ce2} cannot be employed because the prerequisite of comparing two points from the same source position is not fulfilled. Taking into account the relative source position between the two points in the source plane that are mapped onto $r_{A1}$ and $r_{A2}$ in arc $A$ and $r_{B2}$ and $r_{B1}$ in arc $B$, respectively,
\begin{align}
\dfrac{r_{A1} - r_{A2}}{r_{B2} - r_{B1}} =  \dfrac{(1-\kappa_B)}{(1-\kappa_A)} 
\label{eq:ntlo1}
\end{align}
can be derived as detailed in Appendix~\ref{app:ntlo}. Thus, the ratio of arc widths is linked with the ratio of convergences consisting of the common axisymmetric part plus the perturbing part at the centres of $A$ and $B$. Comparing this result with that of \cite{bib:Tessore}, it is not surprising, as Equation~\eqref{eq:ntlo1} amounts to mapping the width of one arc onto the width of the other. As shown in \cite{bib:Wagner2}, both sides of Equation~\eqref{eq:ntlo1} become one for vanishing perturbations.

\subsubsection{Summary of theoretical results}
\label{sec:theoretical_results}

\com{Figure~\ref{fig:configurations} qualitatively shows the different types of multiple image configurations that arise from lenses with perturbed axisymmetry. Deriving the respective equations for each case from the general Equations~\eqref{eq:ce1} and \eqref{eq:ce2}, the observable image properties as stated in Section~\ref{sec:observables} constrain different combinations of (differences in) the components of the perturbing deflection angle and the convergence of the lens at the Einstein radius. }

\com{If external shear is perturbing the main lens in Equation~\eqref{eq:ce1}, the perturber does not contribute to the convergence and $\kappa_\mathrm{a}(r_\mathrm{E})$ is the total convergence at the Einstein radius.}
\com{Without inserting a lens model, no physical properties of the lens can be derived, for example\ the position of the perturber. Furthermore, all sets of equations in Sections~\ref{sec:case_b}, \ref{sec:case_c}, and \ref{sec:case_d} are underconstrained, so that only ratios of the unknown lens properties can be determined. This gives rise to model-independent degeneracies which are investigated in detail in Section~\ref{sec:degeneracies}.}

\section{Limits}
\label{sec:limits}

\subsection{Topologically invariant critical curve}

In \cite{bib:Wagner1}, I showed that the influence of the source properties is negligible in the vicinity of a critical curve, which also applies to the multiple image configurations considered here.

The approach breaks down if $\left| \psi_\mathrm{p}(r,\theta) \right| \ll \left| \psi_\mathrm{a}(r,\theta) \right|$ is no longer valid  and perturbers change the multiplicity of observed images, implying a topological change in the shape of the critical curves. This is the case if a point mass perturber splits an arc into two segments or if a galaxy-scale perturber introduces an ellipticity of the total lensing potential to split the two giant arcs into a four-image configuration typical of elliptical lensing potentials. Formulated in the source plane, the caustic must remain close to point-like compared to the scale of the source size. 

Without any perturbation, the critical curve of an axisymmetric potential producing two tangential arcs is a circle of radius $r_\mathrm{E}$. Using the Taylor expansion in Section~\ref{sec:theory}, to leading order the critical curve is perturbed into a curve parametrised by
\begin{equation}
2(1 - \kappa_\mathrm{a}(r_\mathrm{E})) r^2 -\left( 2 r_\mathrm{E}(1-\kappa_\mathrm{a}(r_\mathrm{E})) + p_1(\theta) \right) r - \dfrac{\partial^2 p_0(\theta)}{\partial \theta^2} = 0\;,
\label{eq:perturbed_cc}
\end{equation}
as derived in Appendix~\ref{app:perturbed_cc}. This is a quadratic equation with two solutions, yielding a tangential and a radial critical curve. In order to shrink the radial critical curve to zero, the last term on the left-hand side must be much smaller than the other two terms. 
Hence, the solution that yields one perturbed critical curve, i.e.\ that leaves the topology of the critical curves invariant, is obtained if
\begin{equation}
\left| \dfrac{\partial^2 p_0(\theta)}{\partial \theta^2} \right| \ll  \left| 2r_\mathrm{E}^2(1-\kappa_\mathrm{a}(r_\mathrm{E})) \right| \; \wedge \; \left|  p_1(\theta) \right| \ll \left| 2r_\mathrm{E}(1-\kappa_\mathrm{a}(r_\mathrm{E})) \right|
\label{eq:limit_conditions}
\end{equation}
is fulfilled. Then, the critical radius $\tilde{r}_\mathrm{E}$, now $\theta$-dependent due to the perturber, is given by
\begin{equation}
\tilde{r}_\mathrm{E}(\theta) = r_\mathrm{E} + \dfrac{p_1(\theta)}{2(1 - \kappa_\mathrm{a}(r_\mathrm{E}))}\;.
\label{eq:perturbed_r_E}
\end{equation}
\com{Practically, a change in the topology of the critical curve due to the perturber becomes visible when the giant arc $A$ is split into three multiple images. Equation~\eqref{eq:perturber_c2} states that this is the case as soon as the derivatives of the perturbing potential in angular direction at the positions of the multiple images differ. Then, the approach developed in \cite{bib:Wagner1} can be used to extract the model-independent properties from those images.}

\subsection{Impact of the perturber on the lensing mass}

The impact on the lensing mass enclosed by the perturbed Einstein radius is estimated by expressing $\tilde{r}_\mathrm{E}(\theta)$ as an effective, perturbed Einstein radius $r_\mathrm{E} + \delta r$, such that the projected lensing mass amounts to
\begin{align}
m(r_\mathrm{E}+\delta r) 
&= \int \limits_0^{2\pi} \mathrm{d} \theta \int \limits_0^{r_\mathrm{E} + \delta r} \mathrm{d}r \; r \left( \kappa_\mathrm{a}(r) + \kappa_\mathrm{p}(r,\theta) \right) \\
&= m_\mathrm{a}(r_\mathrm{E}) + \left(m_\mathrm{a}(r_\mathrm{E} + \delta r) - m_\mathrm{a}(r_\mathrm{E}) \right) + m_\mathrm{p}(r_\mathrm{E} + \delta r) \;.
\label{eq:lensing_mass}
\end{align}
The first term is the unperturbed mass, the second the change of mass in the axisymmetric lens due to the perturbed Einstein radius, and the last term is the mass of the perturber.

\subsection{External shear as an example for $\psi_\mathrm{p}$}

\com{Due to the axial symmetry of the main lens, the external shear can be aligned with the coordinate axes of the lens without loss of generality. Then, the perturbing deflection potential $\psi_\mathrm{p}$ centred at the origin of the coordinate system can be expressed as }
\begin{equation}
\psi_\mathrm{p} (r,\theta) = \dfrac12 \Gamma_1 r^2 \cos(2\theta) \;,
\label{eq:external_shear}
\end{equation}
\com{with the amplitude of the external shear $\Gamma_1$. According to Equation~\eqref{eq:coefficients}, Equation~\eqref{eq:perturbed_cc} reads}
\begin{equation}
\tilde{r}_\mathrm{E}(\theta) = r_\mathrm{E} \left( 1 + \tilde{\Gamma}_1 \cos(2\theta) \right) \;, \quad \tilde{\Gamma}_1 = \dfrac{\Gamma_1}{2(1-\kappa_\mathrm{a}(r_\mathrm{E}))} \;.
\end{equation}
\com{Rewriting the term in brackets yields}
\begin{equation}
\tilde{r}_\mathrm{E}(\theta)= r_\mathrm{E} \left( (1+\tilde{\Gamma}_1) \cos^2(\theta) + (1-\tilde{\Gamma}_1) \sin^2 (\theta) \right) \;,
\end{equation}
\com{which parametrises an ellipse, as expected from Section~\ref{sec:theory}.}

\com{Equation~\eqref{eq:limit_conditions} reads for the perturbing potential of Equation~\eqref{eq:external_shear}}
\begin{equation}
\Gamma_1 \ll 2\left| 1 - \kappa_\mathrm{a}(r_\mathrm{E}) \right| \; \wedge \; \Gamma_1 \ll \left| 1 - \kappa_\mathrm{a}(r_\mathrm{E}) \right| \;,
\end{equation}
\com{thus for $\Gamma_1 \ll \left| 1 - \kappa_\mathrm{a}(r_\mathrm{E}) \right|$ the topology of the critical curve is not changed.}

\com{Taking into account that external shear does not contribute to the mass inside the Einstein radius, the last term in Equation~\eqref{eq:lensing_mass} vanishes for this case. Assuming that the change in mass due to the second term can be Taylor expanded, Equation~\eqref{eq:lensing_mass} reads}
\begin{align}
m(\tilde{r}_\mathrm{E}) = m_\mathrm{a}(r_\mathrm{E}) + \left. \dfrac{\partial m}{\partial r} \right|_{r=r_\mathrm{E}} \delta r =  m_\mathrm{a}(r_\mathrm{E})  + \dfrac{\Gamma_1 \pi r_\mathrm{E}^2 \kappa_\mathrm{a}(r_\mathrm{E})}{1 - \kappa_\mathrm{a}(r_\mathrm{E})} \;.
\end{align}
\com{Hence, the perturbation in the total lensing mass scales with $\Gamma_1 r_{E}^2$ in this case.}

%

\section{Degeneracies}
\label{sec:degeneracies}

\com{So far, the following degeneracies have been found for perturbed Einstein rings:
\begin{itemize}
\item Source position transformation (SPT) \citep{bib:Schneider};
\item Mass sheet degeneracy (MSD), which is a special case of the SPT for a constant mass sheet that can be added to the lens leaving all observables invariant \citep{bib:Falco};
\item Degeneracies between different lens models (DLM), as analysed in \cite{bib:Sluse} and \cite{bib:Vegetti2014},
\item Degeneracies between model parameters (DMP), as detailed in \cite{bib:Vegetti2014}.
\end{itemize}
Using the model-independent Equations~\eqref{eq:ce1}, \eqref{eq:ce2}, and \eqref{eq:ntlo1}, I investigate the connection between the first three degeneracies under the assumption that the multiple images are close enough to the critical curve such that their observables are dominated by the lensing effect\footnote{For the general case, \cite{bib:Wagner1} showed the regions around the critical curve of simulated lenses for which this assumption is a good approximation, assuming that a deviation of 10\% from the true value is considered a tolerable inaccuracy.}.
}

\com{First, I analyse which transformations of the right-hand side leave the left-hand side of Equation~\eqref{eq:ce1} invariant. If the difference of the radial deflection angles in the numerator is transformed by an arbitrary function $f$,
\begin{equation}
\Delta \alpha \equiv \alpha_{\mathrm{p}, r}(r_\mathrm{E}, \theta_i) - \alpha_{\mathrm{p}, r}(r_\mathrm{E}, \theta_j) \; \rightarrow \; \Delta \hat{\alpha} = f(\Delta \alpha) \;,
\end{equation}
and Equation~\eqref{eq:ce1} remains invariant under this transformation, $\kappa_\mathrm{a}$ also has to be transformed to $\hat{\kappa}_\mathrm{a}$. This is determined by solving
\begin{equation}
\Delta r = \dfrac{\Delta \alpha}{2(1-\kappa_\mathrm{a})} =  \dfrac{\Delta \hat{\alpha}}{2(1-\hat{\kappa}_\mathrm{a})} 
\label{eq:radial_degeneracy}
\end{equation}
for $\hat{\kappa}_\mathrm{a}$ to obtain the transformation
\begin{equation}
\hat{\kappa}_\mathrm{a} = \lambda_\alpha \kappa_\mathrm{a} + (1 - \lambda_\alpha) \;, \quad \lambda_\alpha = \dfrac{\Delta \hat{\alpha}}{\Delta \alpha} \;.
\label{eq:kappa_transformation}
\end{equation}
Vice versa, if $\kappa_\mathrm{a}$ is transformed by an arbitrary function $\hat{\kappa}_\mathrm{a} = f(\kappa_\mathrm{a})$, inserting it in Equation~\eqref{eq:radial_degeneracy}, the resulting pair of transformations reads
\begin{equation}
\hat{\kappa}_\mathrm{a} = f(\kappa_\mathrm{a}) \; \rightarrow \; \Delta \hat{\alpha} = \lambda_\kappa \Delta \alpha \;, \quad \lambda_\kappa = \dfrac{1 - \hat{\kappa}_\mathrm{a}}{1 - \kappa_\mathrm{a}} \;.
\end{equation}
The transformation of Equation~\eqref{eq:kappa_transformation} is the transformation of an MSD. Hence, changing the difference of the radial deflection angles of the perturber can always be compensated for by adding a mass sheet to the main lens. In turn, changing the convergence of the main lens can be balanced by a scaling of $\Delta \alpha$. As the main lens is assumed to be axisymmetric, $f(\kappa_\mathrm{a})$ is a special SPT as detailed in \cite{bib:Schneider}\footnote{Analogously to the MSD, this SPT is an exact invariance transformation that can be derived from a lensing potential. According to \cite{bib:Schneider}, constraints on a physically meaningful convergence are that $f$ is even, positive, and monotonically decreasing.}. Thus, there is a fundamental degeneracy between the main lens and the perturber that can only be broken if their relative positions are known.}

\com{Equation~\eqref{eq:ce2} does not show this degeneracy as it is only dependent on the difference of the angular deflection angles of the perturber. If $\alpha_{\mathrm{p}, \theta}(r_\mathrm{E}, \theta_i)\equiv \alpha_{\mathrm{p}, i}$ is transformed by an arbitrary function $f$, then the transformation function $g$ of $\alpha_{\mathrm{p}, \theta}(r_\mathrm{E}, \theta_j)\equiv \alpha_{\mathrm{p},j}$ determined from
\begin{equation}
\Delta \theta = \dfrac{\alpha_{\mathrm{p}, i}}{r_i\color{black}{^2}} - \dfrac{\alpha_{\mathrm{p}, j}}{r_j\color{black}{^2}} = \dfrac{f(\alpha_{\mathrm{p}, i})}{r_i\color{black}{^2}} - \dfrac{g(\alpha_{\mathrm{p}, j})}{r_j\color{black}{^2}}
\label{eq:angular_degeneracy}
\end{equation}
is given by
\begin{equation}
g(\alpha_{\mathrm{p}, j}) = \alpha_{\mathrm{p}, j} + \dfrac{r_j\color{black}{^2}}{r_i\color{black}{^2}} \left( f(\alpha_{\mathrm{p}, i}) - \alpha_{\mathrm{p}, i}\right) \;.
\label{eq:angular_transformation}
\end{equation}
The angular deflection angle is thus determined up to a constant and up to its own MSD, not knowing its position with respect to the source and the observer.}

\com{Repeating the same procedure for Equation~\eqref{eq:ntlo1}, I set up the equation
\begin{equation}
\dfrac{\Delta r_A}{\Delta r_B} = \dfrac{1-\kappa_B}{1-\kappa_A} =  \dfrac{1-\hat{\kappa}_B}{1-\hat{\kappa}_A} \;, \quad \hat{\kappa_A} = f(\kappa_A) \;, \quad \hat{\kappa_B} = g(\kappa_B) \;,
\label{eq:kappa_ratio_degeneracy}
\end{equation}
that employs the transformation functions $f$ and $g$ of $\kappa_A$ and $\kappa_B$, respectively.
Then, the transformation of $\kappa_B$ amounts to the transformation of an MSD of the transformed $\kappa_A$. 
\begin{equation}
\hat{\kappa}_B = \lambda_{AB} \hat{\kappa}_A + (1-\lambda_{AB}) \;, \quad \lambda_{AB} = \dfrac{1-\kappa_B}{1-\kappa_A} \;.
\label{eq:kappa_ratio_transformation}
\end{equation}
Subsequently, $\hat{\kappa}_A$ can be an SPT. Although $\kappa_A$ is not necessarily axisymmetric, the SPT is restricted to an exact invariance transformation, as already employed above. The most general form of an SPT cannot occur by construction, due to the commuting derivatives of the potential in Equation~\eqref{eq:Taylor_potential}.}

\subsection{Comparison to existing results}

\com{With Eqs.~\eqref{eq:radial_degeneracy} to \eqref{eq:kappa_ratio_transformation}, the results found in \cite{bib:Sluse} can be explained in a model-independent way: using simulations, they found a substantial freedom in the choice of possible lens models that described a given set of multiple images equally well. As models, they employed a power-law profile with external shear and a composite model consisting of a Hernquist profile, a Navarro-Frenck-White (NFW) profile (see Table~\ref{tab:lens_models} for further details), and external shear. The last model is a composite of three parts, but the first two parts add up to an axisymmetric lens so that the external shear can be treated as the perturbation. Eq.~\eqref{eq:kappa_transformation} explains that fixing the MSD for the main lens in both models by a measurement of the velocity dispersion along the line of sight does not break the degeneracy. The measurement fixes the convergence values for both models, i.e. $\hat{\kappa_a}$ and $\kappa_a$. Subsequently, scaling the values for the external shear (perturbation) according to Eq.~\eqref{eq:kappa_transformation} means that both models can yield the same observables.}

\com{Analogously, Eqs.~\eqref{eq:radial_degeneracy} to \eqref{eq:kappa_ratio_transformation} support the results of \cite{bib:Minor} and \cite{bib:Vegetti2014} that the properties of the perturber cannot be determined using the constraints from the observable deviations from an Einstein ring alone and that employing lens models cannot break this degeneracy.}

\subsection{Breaking the degeneracies}

\com{As already pointed out in Section~\ref{sec:degeneracies}, the degeneracy in Equation~\eqref{eq:radial_degeneracy} can only be broken  if the relative position of the main lens and the perturber is known. The same applies to Equation~\ref{eq:kappa_ratio_degeneracy}. Equation~\ref{eq:angular_degeneracy} requires the position of the perturber with respect to the source and the observer to be known. Thus, knowing the position of the main lens and the perturber with respect to the source and the observer breaks all degeneracies. Yet, this requires knowing about the existence of a perturbation in the first place and  its rough properties, for instance whether the perturbation is an external shear caused by the environment or whether it is a black hole or another compact, massive object.}

\begin{table}[ht]
\caption{Parameters chosen in the generalised axisymmetric mass density profile, Eq.~\eqref{eq:parametric_model}, to retrieve some of the most common parametric lens models.}
\begin{tabular}{lllll}
\hline
 \noalign{\smallskip}
     Model & Reference & $\alpha$ & $\beta$ & $\gamma$ \\
 \noalign{\smallskip}
\hline
 \noalign{\smallskip}
Plummer model & (1) & 0 & 2 & 5/2 \\
Non-singular isothermal sphere & (2) & 0 & 2 & 1 \smallskip \\
Navarro-Frenk-White profile & (3) & 1 & 1 & 2 \\
Hernquist model & (4) & 1 & 1 & 3 \smallskip \\
Singular isothermal sphere (SIS) & (2) & 2 & - & 0 \\
Jaffe model & (5) & 2 & 1 & 2 \\
 \noalign{\smallskip}
\hline
\end{tabular}
\tablebib{(1)~\citet{bib:Plummer};
(2)~\citet{bib:SEF};
(3)~\cite{bib:NFW2};
(4)~\cite{bib:Hernquist};
(5)~\cite{bib:Jaffe}.}
\label{tab:lens_models}
\end{table}

The model-independent equations do not make any assumptions about these properties. Therefore, inserting lens models that specify the mass density profiles of the main lens and the perturber into Equations~\eqref{eq:radial_degeneracy} to \eqref{eq:kappa_ratio_transformation}, it becomes possible to estimate possible positions of the perturber and its mass and determine \newpage \noindent the model parameters of the main lens. Thus, if there is observational evidence in favour of a certain model configuration and the positions of the main lens and the perturber along the line of sight between observer and source \comm{are known}, the Bayesian approach developed in \cite{bib:Koopmans} and \cite{bib:Vegetti2009} determines the most likely a posteriori model parameters.

If the galaxy whose multiple images are observed is the host of a quasar, the time delay between the quasar images can be measured. To leading order, the time delay $\Delta t_{AB}$ between the quasar image at position $\boldsymbol{x}_A = (r_A, \theta_A)$ in the arc with positive parity and the quasar image at position  $\boldsymbol{x}_B = (r_B, \theta_B)$ in the arc with negative parity is given by
\begin{equation}
\color{black}
\Delta t_{AB} = \dfrac{\Gamma_\mathrm{d}}{c} \left( \Delta \phi_\mathrm{a} - \delta \boldsymbol{y} \left(\boldsymbol{x}_A - \boldsymbol{x}_B \right) - \Delta \psi_\mathrm{p} \right)
\label{eq:time_delay}
\end{equation}
with the time delay due to the main lens $\Gamma_\mathrm{d}/c \Delta \phi_\mathrm{a}$, the shift in the source position due to the perturber $\delta \boldsymbol{y}$, $\Delta \psi_\mathrm{p} = \psi_\mathrm{p}(\boldsymbol{x}_A) - \psi_\mathrm{p}(\boldsymbol{x}_B)$, and $\Gamma_\mathrm{d} = D_\mathrm{d}D_\mathrm{s}/(D_\mathrm{ds}) (1+z_\mathrm{d})$.
The derivation is given in Appendix~\ref{app:time_delay}. Since quasars are usually brighter than their host galaxy, I assume that the position of the quasar image is also the centre of light of the arc. Hence, measuring time delays can help to break the degeneracy between the main lens and the perturber if the last two terms are small compared to the first one, so that the time delay depends on the main potential only.

\section{Models}
\label{sec:models}

\subsection{Axisymmetric model class}
\label{sec:axisymmetric_model_class}

The most commonly used axisymmetric three-dimensional mass density functions can be described by
\begin{equation}
\rho (R, \alpha, \beta, \gamma) = \dfrac{\rho_0}{R^\alpha \left(1+R^\beta \right)^\gamma} \;,
\label{eq:parametric_model}
\end{equation}
in which $\rho_0$ denotes the density at $R = 0$ and $R$ is the three-dimensional distance to the centre scaled by a typical scale radius $R_\mathrm{s}$. The parameter $\alpha$ determines the slope of the density profile for $R \ll 1$ in the central part, $\beta$ in the intermediate region ($R \approx 1$), and $\gamma$ in the outer parts ($R \gg 1$). The values for $\alpha, \beta, \gamma  \in \mathbb{R}$ of the most common lens models are listed in Table~\ref{tab:lens_models}. 

Projecting the three-dimensional mass density profile along the line of sight, the convergence is obtained as
\begin{align}
\kappa_\mathrm{a}(r,\alpha, \beta, \gamma) = \dfrac{2 \kappa_0}{r^{\alpha-1}} \int \limits_0^{\pi/2} \mathrm{d} \chi \dfrac{(\sin(\chi))^{\alpha+\beta\gamma-2}}{\left( (\sin(\chi))^\beta + r^\beta \right)^\gamma} \le \dfrac{\pi \kappa_0 r_\mathrm{E}^{1-\alpha}}{\left(1 + r_\mathrm{E}^\beta \right)^\gamma} \;,
\end{align}
where $r$ denotes the two-dimensional radius and $\chi$ the integration variable along the line of sight. As $\kappa_\mathrm{a}(r,\alpha,\beta,\gamma)$ is bounded from above, there is a maximum scaling of Eq.~\eqref{eq:ce1}.
For the case of a singular isothermal sphere (SIS), $\kappa_\mathrm{a} = 1/2$, such that Eq.~\eqref{eq:perturbed_r_E} becomes 
\begin{equation}
\tilde{r}_\mathrm{E}(\theta) = r_\mathrm{E} + p_1(\theta) \;.
\end{equation}

\subsection{Point mass perturber}
\label{sec:point_mass_perturber}

In this section and the following one, all angular coordinates $\theta$ are in $\left[ 0, 2\pi \right]$, i.e.\ the deflection angle of the perturber is determined in the coordinate system in which the centre of the axisymmetric potential is the origin.

Specifying the perturber as point mass, e.g.\ a transient or a black hole, at $\boldsymbol{x}_\mathrm {p} = (r_\mathrm{p}, \theta_\mathrm{p})$, the deflection potential of the perturber reads
\begin{equation}
\psi_\mathrm{p}(\boldsymbol{x}) = r_\mathrm{E,p}^2 \ln(||\boldsymbol{x}-\boldsymbol{x}_\mathrm{p}||) \;,
\end{equation}
where $r_\mathrm{E,p}$ is the Einstein radius of the point mass. Deriving the components of the deflection angle in polar coordinates yields
\begin{align}
\alpha_{\mathrm{p},r}(r,\theta) &= \dfrac{r_\mathrm{E,p}^2  \left( r - r_\mathrm{p} \cos (\theta - \theta_\mathrm{p})\right) }{r^2 + r_\mathrm{p}^2 - 2 r r_\mathrm{p}\cos (\theta - \theta_\mathrm{p})} \;,\label{eq:pmp1} \\
\alpha_{\mathrm{p},\theta}(r,\theta) &=  \dfrac{r_\mathrm{E,p}^2  r_\mathrm{p} \color{black}{r} \color{black} \sin (\theta - \theta_\mathrm{p})}{r^2 + r_\mathrm{p}^2 - 2 r r_\mathrm{p}\cos (\theta - \theta_\mathrm{p})} \;. \label{eq:pmp2}
\end{align}

\begin{figure*}[ht]
\centering
\begin{subfigure}{0.2\textwidth}
  \centering
  \hspace{-7ex}
 \includegraphics[width=0.6\linewidth]{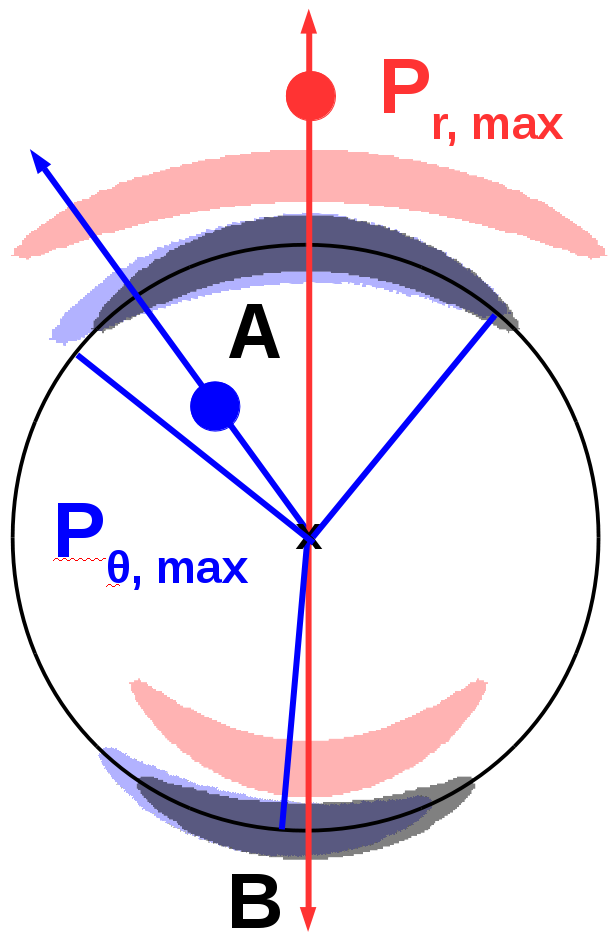}
 \caption{}
 \end{subfigure}
\hspace{-6ex}
 \begin{subfigure}{0.27\textwidth}
  \centering
 \includegraphics[width=0.95\linewidth]{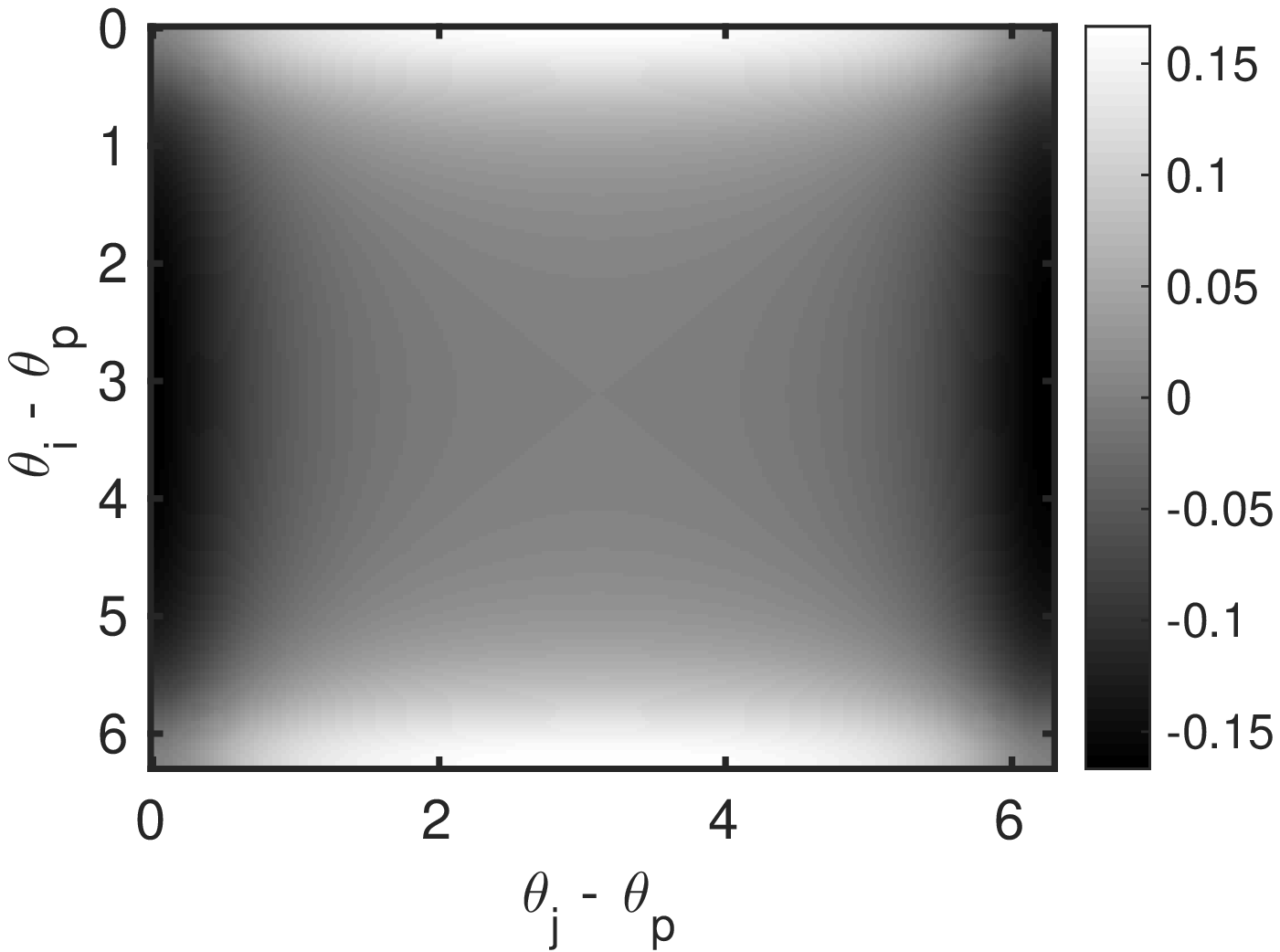}
  \caption{}
 \end{subfigure}
\hspace{0.4ex}
 \begin{subfigure}{0.27\textwidth}
  \centering
 \includegraphics[width=0.95\linewidth]{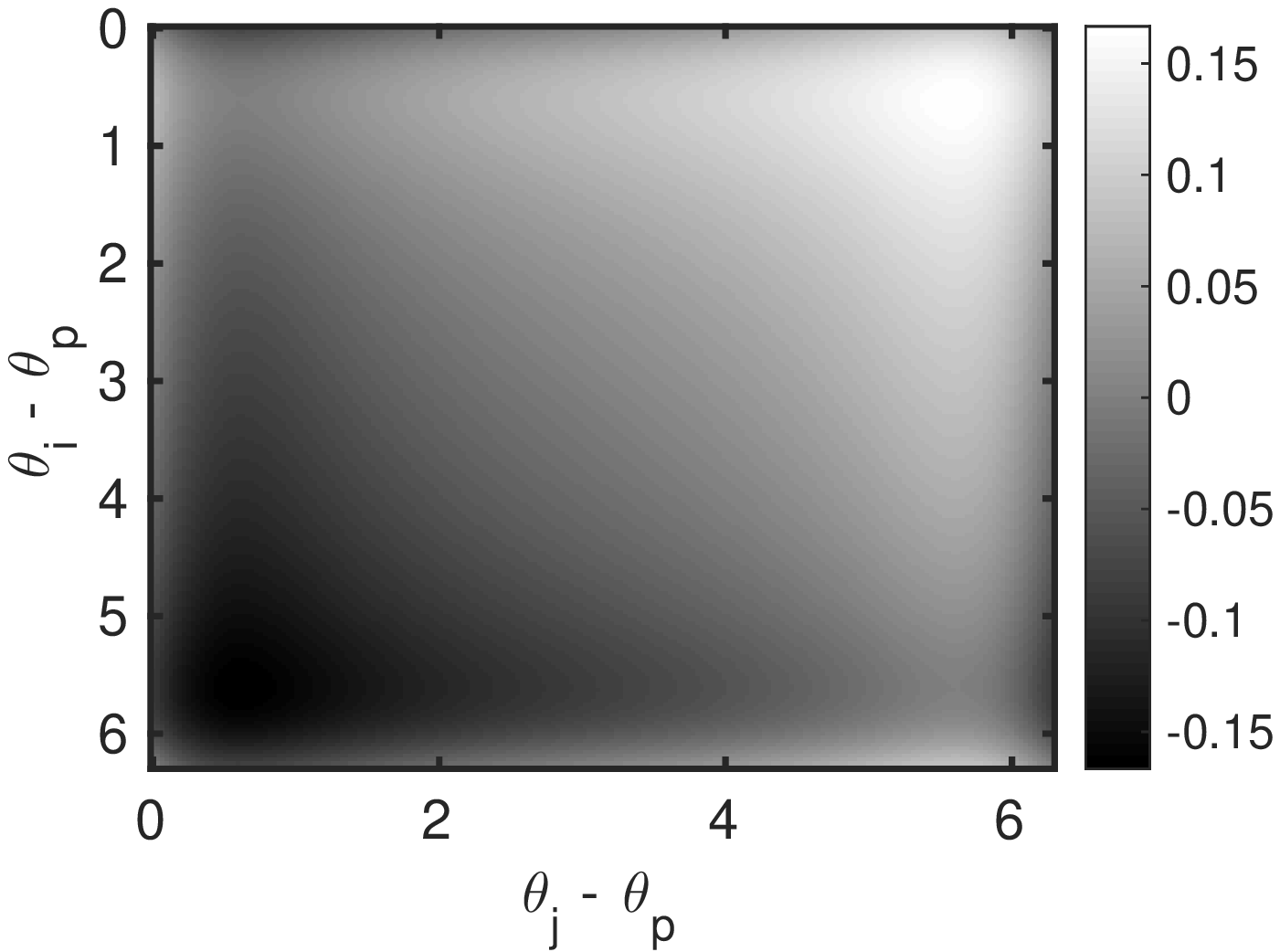}
  \caption{}
 \end{subfigure}
 \hspace{0.4ex}
  \begin{subfigure}{0.27\textwidth}
  \centering
 \includegraphics[width=0.93\linewidth]{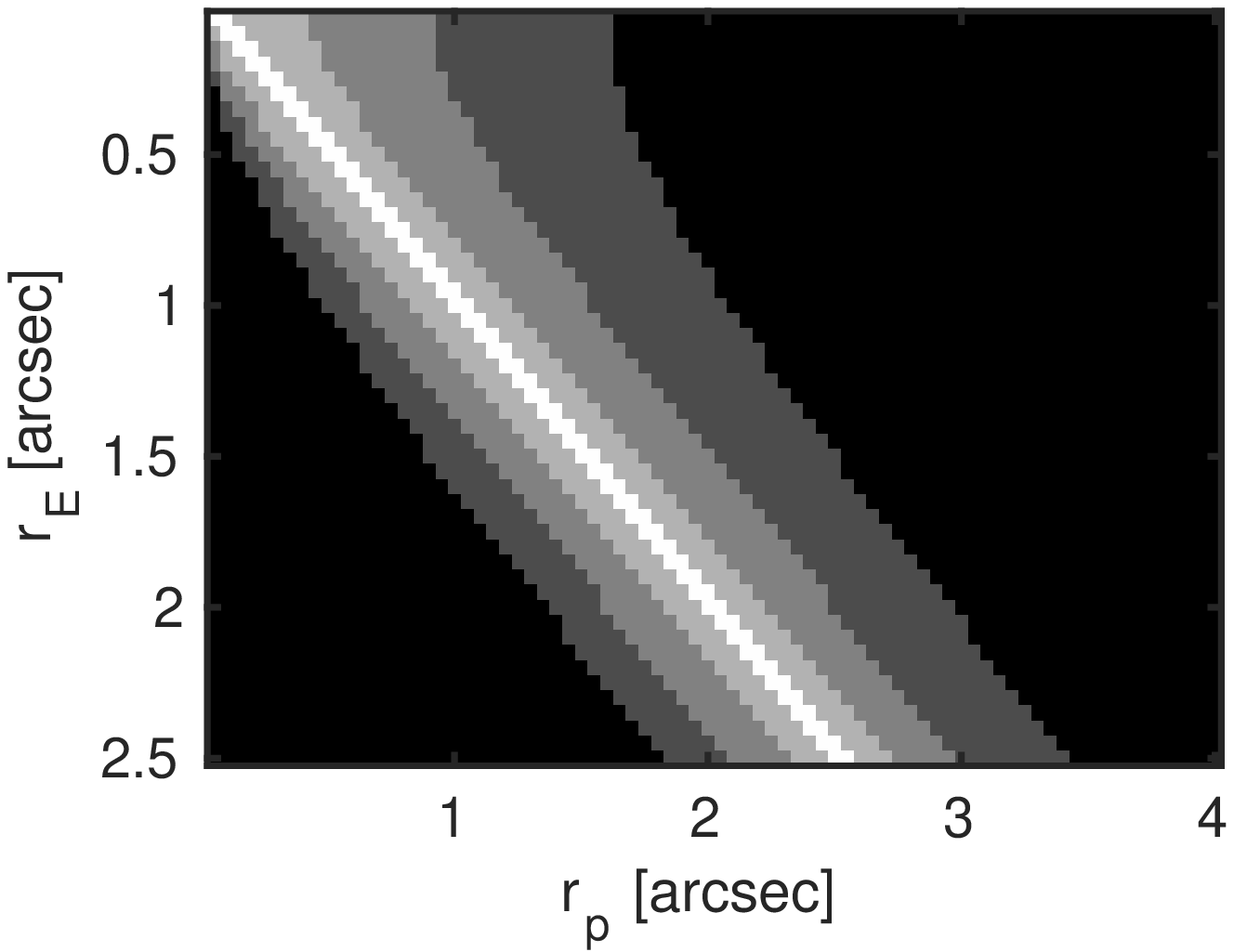}
  \caption{}
 \end{subfigure}
    \caption{Configurations of point mass perturbers causing a maximum difference in radii (red) or a maximum difference in arc lengths of the individual arcs (blue), excluding the origin point (a); Eq.~\eqref{eq:sis_pmp1} (b) and Eq.  \eqref{eq:sis_pmp2} (c) with respect to\ the alignments of the images and the perturber using $r_\mathrm{E} = 2.0''$, $r_\mathrm{E,p} = 0.5''$, and $r_\mathrm{p} = 1.0''$; Equation~\eqref{eq:max_diff_pm} above $s_0 = 0.05''$ varying the Einstein radius of the point mass from $0.2'', 0.15'', 0.1''$ to $0.05''$ (dark grey to white shaded area) (d).}
    \label{fig:pmp}
\end{figure*}

\subsection{Singular isothermal sphere galaxy-scale perturber}

Assuming the perturber is a galaxy located at  $\boldsymbol{x}_\mathrm {p} = (r_\mathrm{p}, \theta_\mathrm{p})$, the deflection potential of the perturber reads
\begin{equation}
\psi_\mathrm{p}(\boldsymbol{x}) =  r_\mathrm{E,p} || \boldsymbol{x} - \boldsymbol{x}_0 || \;
\end{equation}
if the galaxy mass density profile can be approximated by a singular isothermal sphere, which is also employed as an appropriate galaxy-scale substructure model in \cite{bib:Vegetti2014}.
Then, the components of the deflection angle are given by 
\begin{align}
\alpha_{\mathrm{p},r}(r,\theta) &= \dfrac{r_\mathrm{E,p}  \left( r - r_\mathrm{p} \cos (\theta - \theta_\mathrm{p})\right) }{\sqrt{r^2 + r_\mathrm{p}^2 - 2 r r_\mathrm{p}\cos (\theta - \theta_\mathrm{p})}} \;,\label{eq:sisp1} \\
\alpha_{\mathrm{p},\theta}(r,\theta) &=  \dfrac{r_\mathrm{E,p}  r_\mathrm{p} \color{black}{r} \color{black} \sin (\theta - \theta_\mathrm{p})}{\sqrt{r^2 + r_\mathrm{p}^2 - 2 r r_\mathrm{p}\cos (\theta - \theta_\mathrm{p})}} \;. \label{eq:sisp2}
\end{align}

\section{Detectable scales of perturbers}
\label{sec:substructure}

\subsection{General estimates}
\label{sec:general_estimates}

Given the minimum observable length scale $s_\mathrm{o}$, which is constrained by the resolution and the signal-to-noise ratio in the data, the radial difference in Equation~\eqref{eq:ce1} must be larger than $s_\mathrm{o}$ to be observable. Analogously, I reformulate the left-hand side of Equation~\eqref{eq:ce2}
\begin{align}
r_j (r_i \theta_i) - r_i (r_j \theta_j) \equiv r_j a_i - r_i a_j \approx r_\mathrm{E} \left( a_i - a_j \right) \;, 
\end{align}
\newpage \noindent
denoting the arc length from the centre of the arc to the angle $\theta$ by $a$ and approximating $r_i \approx r_j \approx r_\mathrm{E}$. Hence, 
\begin{align}
r_i - r_j &= \dfrac{\alpha_{\mathrm{p}, r}(r_\mathrm{E}, \theta_i) -  \alpha_{\mathrm{p}, r}(r_\mathrm{E}, \theta_j)}{2(1-\kappa_\mathrm{a}(r_\mathrm{E}))} > s_\mathrm{o} \;, \label{eq:observable_perturbation1} \\
a_i - a_j &= \color{black}{\tfrac{\alpha_{\mathrm{p}, \theta}(r_\mathrm{E}, \theta_i) -  \alpha_{\mathrm{p}, \theta}(r_\mathrm{E}, \theta_j)}{r_\mathrm{E}}} \textcolor{black}{ > s_\mathrm{o}} \label{eq:observable_perturbation2}
\end{align}
give a lower bound of observable properties of perturbers. 

An optimistic estimate $s_\mathrm{o}$ of ground-based observations is given by $0.5''$  \ \cite[e.g.][]{bib:Erben} and $0.05''$ for space-based (HST-like) observations. As the distribution of Einstein radii of galaxy-scale lenses lies in the range of $0.5''$ to about $2''$ \ \cite[e.g.][]{bib:Bolton}, detecting perturbers in ground-based observational data via the approach proposed here is hardly possible as the Einstein radii of the lenses are already of the order of $s_\mathrm{o}$. For space-based observations, $s_\mathrm{o}$ is improved by up to one order of magnitude and the influence of perturbers becomes measurable. However, using adaptive optics for ground-based observations may yield better $s_0$ than the $s_0$ obtained in low-resolution and low signal-to-noise HST images, so that adaptive optics can lead to more detailed observations, as investigated on B1938+666 in \cite{bib:Lagattuta}. 

In the following I will show the examples of a point mass perturber and a galaxy-scale perturber to estimate the detectable properties of both kinds of perturbers \comm{under the assumption that the degeneracies detailed in Sect.~\ref{sec:degeneracies} could be resolved.} As \comm{an} axisymmetric lens that generates the Einstein ring, I will assume a singular isothermal sphere with $\kappa_\mathrm{a} = 1/2$ as introduced in Sect.~\ref{sec:axisymmetric_model_class}.

\comm{In Sect.~\ref{sec:comparison}, I show that these estimates are consistent with those obtained by simulations  \cite[e.g.][]{bib:Vegetti2009, bib:Vegetti2014}, and other ansatzes to describe perturbers,  \cite[e.g.][]{bib:Minor}. Thus, coupling the model-independent equations, Eqs.~\eqref{eq:ce1} and \eqref{eq:ce2}, with models enables us to derive analytic constraints on the properties of the perturber and their degeneracies}. 

\subsection{Point mass perturbers to a singular isothermal sphere}
\label{sec:point_mass_substructure}

Inserting Equations~\eqref{eq:pmp1} and \eqref{eq:pmp2} into the right-hand sides of Equations~\eqref{eq:observable_perturbation1} and \eqref{eq:observable_perturbation2} yields
\begin{align}
r_i - r_j &= \dfrac{r_\mathrm{E,p}^2 r_\mathrm{p}\left( r_\mathrm{E}^2 - r_\mathrm{p}^2\right) \left( \cos(\theta_i - \theta_\mathrm{p}) - \cos(\theta_j - \theta_\mathrm{p})\right)}{D(\theta_i) D(\theta_j)} \label{eq:sis_pmp1} \;, \\
a_i - a_j &=  \dfrac{r_\mathrm{E,p}^2  r_\mathrm{p} \left( D(\theta_j) \sin (\theta_i - \theta_\mathrm{p}) - D(\theta_i) \sin (\theta_j - \theta_\mathrm{p}) \right)}{D(\theta_i) D(\theta_j)} \label{eq:sis_pmp2} \;,
\end{align}
with
\begin{equation}
D(\theta) =  r_\mathrm{E}^2  + r_\mathrm{p}^2 - 2 r_\mathrm{E} r_\mathrm{p} \cos(\theta - \theta_\mathrm{p}) \;, \quad \theta, \theta_i, \theta_j \in \left[  0, 2\pi\right] \;.
\end{equation}
Hence, the maximum absolute deviation in $r$ due to a perturbing point mass is observed when the images and the perturbing point mass are aligned, and the minimum absolute radial deviation, namely none, is observed when $\theta_\mathrm{p}$ is orthogonal to $\theta_i$ and $\theta_j$ (excluding the origin, for which the images and the perturber are aligned). For the deviation in arc lengths, the maximum absolute deviation is reached for a perturbing point mass located such that
\begin{align}
\theta_i - \theta_\mathrm{p} = \mathrm{acos} \left( \dfrac{2 r_\mathrm{E} r_\mathrm{p}}{r_\mathrm{E}^2 + r_\mathrm{p}^2} \right)\;,  \quad  \theta_j - \theta_\mathrm{p} = -\mathrm{acos} \left( \dfrac{2 r_\mathrm{E} r_\mathrm{p}}{r_\mathrm{E}^2 + r_\mathrm{p}^2} \right)\;,
\end{align}
and the minimum absolute angular deviation, namely none, is reached for a perturbing point mass aligned with both images. Altogether
\begin{align}
\left| r_i - r_j \right| \in \left[0, \dfrac{2 r_\mathrm{E,p}^2 r_\mathrm{p}}{\left| r_\mathrm{E}^2 - r_\mathrm{p}^2 \right|} \right] \;, \quad \left| a_i - a_j \right| \in \left[0, \dfrac{2 r_\mathrm{E,p}^2 r_\mathrm{p}}{\left| r_\mathrm{E}^2 - r_\mathrm{p}^2 \right| } \right] \;. \label{eq:max_diff_pm}
\end{align} 
Figure~\ref{fig:pmp} (a) sketches both cases, and (b) and (c) show Equations~\eqref{eq:sis_pmp1} and \eqref{eq:sis_pmp2} with respect to the alignment of the images and the perturbing mass for $r_\mathrm{E} = 2.0''$, $r_\mathrm{E,p} = 0.5''$, and $r_\mathrm{p} = 1.0''$.

Given Equation~\eqref{eq:max_diff_pm} and $s_0 = 0.05''$, the minimum mass of a detectable point mass perturber can be estimated, assuming that the perturber is aligned to yield maximum differences in the radii or arc lengths. For a perturbing point mass with an Einstein radius in the range of $0.05''$ to $0.2''$, its mass is in the range of $6.1 \cdot 10^8 M_\odot$ to $9.7 \cdot 10^{9} M_\odot$, making the detection of transients improbable.

Figure~\ref{fig:pmp} (d)  shows the ranges of $r_\mathrm{p}$ and $r_\mathrm{E}$ where an optimally positioned perturbing point mass with varying Einstein radius results in observable differences (according to Equation~\eqref{eq:max_diff_pm}) above $s_0$. Decreasing the Einstein radius from $0.2'', 0.15'', 0.1''$ to $0.05''$,  the dark grey to white shaded areas indicate observable positions of a perturbing point mass. Obviously, the range of observable point mass perturbers is limited to the vicinity of the critical curve of the axisymmetric potential.

\subsection{Galaxy-scale perturbers to a singular isothermal sphere}
\label{sec:galaxy_substructure}

While point mass perturbers can be located anywhere, galaxy-scale perturbers are restricted to $r_\mathrm{p} > r_\mathrm{E}$ if no galaxy-merging event is considered. Then, performing the same calculations as in Section~\ref{sec:point_mass_substructure}, I arrive at 
\begin{align}
\left| r_i - r_j \right| \in \left[0, 2 r_\mathrm{E,p} \right] \;, \quad \left| a_i - a_j \right| \in \left[0, 2 r_\mathrm{E,p} \right]  \label{eq:max_diff_sis}
\end{align} 
for the range of maximum absolute radial and angular deviations due to an SIS-perturber, respectively. The maximum difference in radii is reached when the perturber is aligned with the images; for the maximum absolute difference in arc lengths, the positions of the images and the perturber are
\begin{align}
\theta_i - \theta_\mathrm{p} = \mathrm{acos} \left( \dfrac{r_\mathrm{E} }{r_\mathrm{p}} \right) \;, \quad \theta_j - \theta_\mathrm{p} = \mathrm{acos} \left( \dfrac{r_\mathrm{E} }{r_\mathrm{p}} \right) \;,\quad r_\mathrm{E} < r_\mathrm{p} \;.
\end{align}
Aligning the perturbing SIS optimally with the images according to these equations, perturbers with an Einstein radius larger than $0.025''$ and a corresponding mass larger than $9.7 \cdot 10^7 M_\odot$ are observable for $s_0 = 0.05''$.

\subsection{Comparison to existing approaches}
\label{sec:comparison}

Summarising the results obtained in Sections~\ref{sec:point_mass_substructure} and \ref{sec:galaxy_substructure}, I conclude that under an optimal alignment of the perturber and the images, perturbing point masses and SIS of more than $10^{8} M_\odot$ are detectable under HST-like observational conditions if they lie in the near vicinity of the critical curve of the axisymmetric lens. This is in accordance with the results of \cite{bib:Vegetti2009}. Using simulations of elliptical power-law density profiles, they find that masses of $10^{7} M_\odot$ can be detected if they are located on the Einstein ring of the power-law lensing potential, and masses of at least $10^{9} M_\odot$ are required to be observed in the vicinity of the critical curve.

Considering the degeneracies discussed in Section~\ref{sec:degeneracies}, I can also confirm the findings of \cite{bib:Minor} and \cite{bib:Vegetti2014} that the mass density profile and the position of the perturber remain degenerate to a high degree, so that models can only be excluded by the observed image configuration.

\section{Example: B1938+666}
\label{sec:example}
\label{sec:B1938}

\begin{figure}[ht]
  \centering
 \includegraphics[width=0.43\linewidth]{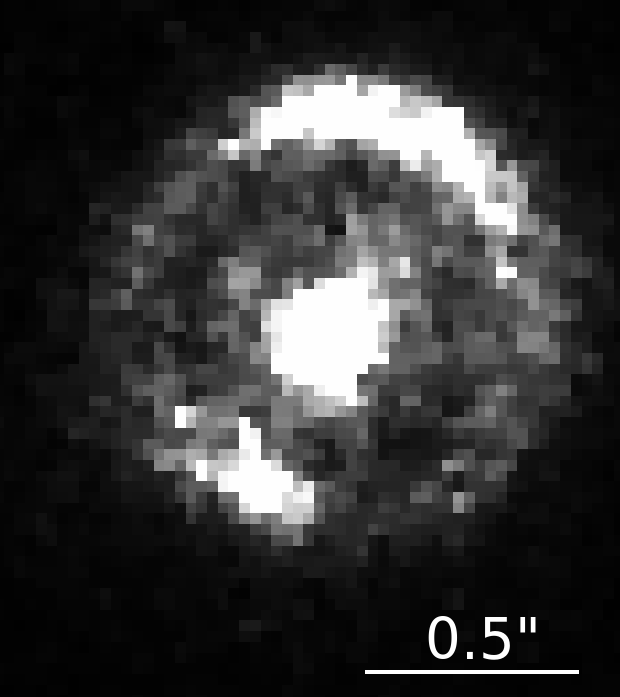}
\caption{HST NIC1 observation of B1938+666 in the F160W band filter (Proposal ID 7255, PI: Jackson, 1997).}
    \label{fig:B1938}
\end{figure}

\comm{To show that existing approaches and the theoretical derivations used to characterise perturbers are also consistent when applied to observational data, I consider the galaxy-scale lens JVAS B1938+666 as an example.}
\comm{It is located} at $z_\mathrm{d} = 0.881$ and has a highly axisymmetric mass distribution, generating an almost perfect Einstein ring of a source galaxy at $z_\mathrm{s} = 2.059$, as shown in Figure~\ref{fig:B1938}. 

Fitting a circle to the Einstein ring, $r_\mathrm{E} = 0.45''$ is obtained. Fitting two circles to the upper and lower arc, denoted as image $A$ and $B$, respectively, I arrive at $r_A = (0.50 \pm 0.1)''$ and $r_B = (0.41 \pm 0.1)''$. Since the signal-to-noise ratio is not high enough to resolve brightness features in the arcs, angular information cannot be extracted and source properties are negligible.

Estimating the scale of a galaxy-scale perturbing mass that causes the difference in the radii by Equation~\eqref{eq:max_diff_sis} yields a mass estimate of
\begin{equation}
m_\mathrm{p} = \left( 5.89 \pm ^{2.91}_{2.30} \right) 10^8 M_\odot
\label{eq:mp_B1938}
\end{equation}
 for an optimally aligned SIS as perturber to an SIS as axisymmetric lens. 

\com{Splitting the potential in an axisymmetric lens with only one perturbing potential and manually extracting the observables may be simplistic. However, the order of magnitude of the perturbing mass in Equation~\eqref{eq:mp_B1938} agrees well with the result $m_\mathrm{p} = (1.9 \pm 0.1) 10^8 M_\odot$ obtained by \cite{bib:Vegetti2012}. Their approach also assumes that the predicted position of the perturber is not far from the critical curve, otherwise a mass of this range could not have been detected. Furthermore, their approach applies a more sophisticated elliptical model for the main lens that additionally includes external shear.} 

\com{However, as pointed out in Sections~\ref{sec:theoretical_results} and \ref{sec:degeneracies}, more sophisticated models and an automated image processing do not necessarily yield more accurate predictions. None of these results breaks the degeneracy between the main lens and the perturber as the relative position between the main lens and the perturber remains unknown. Therefore, as in Section~\ref{sec:comparison}, the goal was to show that the equations derived in Section~\ref{sec:equations} are in agreement with existing approaches and lead to similar results.}

\com{To reduce the degeneracies in this system, a measurement of the velocity dispersion in the lensing potential could be performed.  If the main perturbation is caused by external shear of the environment, the measurement of the velocity dispersion fixes $\kappa_\mathrm{a}$. To determine the impact of the environment and estimate the strength of external shear, the surrounding galaxies have to be characterised, as  investigated in e.g. \cite{bib:McCully} and \cite{bib:Sluse2}.}

\section{Conclusion}
\label{sec:conclusion}

In this work, I employed the same principle as in \cite{bib:Wagner1} and \cite{bib:Wagner2} to derive an analytic, model-independent characterisation of a lens configuration consisting of an axisymmetric main lens and a small perturbation. The resulting equations link observable differences in the arc radii, which almost form an Einstein ring, to differences in the radial component of the deflection angles of the perturber weighted by the convergence of the main lens. \com{The thickness of the arc radii is linked to ratios of convergences determined at the positions of the arcs and is thereby a special case of the more general results developed in \cite{bib:Wagner2}}. Asymmetries in the angular brightness distributions within the arcs are linked to differences in the angular component of the deflection angle of the perturber \comm{for sources with symmetric brightness profiles aligned or orthogonal to the caustic}. In addition, I derive general constraints on the derivatives of the perturbing lens potential that are required to consider the perturber as a small deviation to the main lens, i.e. not changing the number of critical curves and caustics. 

In order to obtain physical properties of the perturber, as its mass or its position, models to determine the convergence of the axisymmetric lensing potential and the deflection angle of the perturber have to be inserted. Then, the approach leads to results that are in agreement with the findings of \cite{bib:Vegetti2009}, \cite{bib:Vegetti2014}, \cite{bib:Minor}, \cite{bib:Sluse2}, and \cite{bib:Vegetti2012} concerning detectable mass scales and the degeneracies due to the model-dependence and the perturbing mass. \com{As proof of principle, I apply the derived equations with a simple configuration of models to B1938+666 and find that the perturbing mass is of the same order of magnitude as the value derived by \cite{bib:Vegetti2012}.}

\com{The values obtained are degenerate, as already found in simulations investigated in \cite{bib:Minor}, \cite{bib:Sluse2}, and \cite{bib:Vegetti2014}. The model-independent approach derived in this work explains these degeneracies from a more general viewpoint, especially the fact that the degeneracy between the perturber and the main lens is not only a degeneracy between different classes of lens models, but a model-independent degeneracy already contained in the general lensing equations. From them, it can be concluded that all degeneracies can be broken if the positions of the lens and the perturber with respect to the observer and the source are known. Measurements of the velocity dispersion along the line of sight and a characterisation of the galaxies in the environment of the lens can contribute to reducing or even breaking the degeneracies, as already successfully applied in \cite{bib:Suyu}. Furthermore, measuring time delays of quasars whose host galaxy is mapped to a perturbed Einstein ring can also contribute to breaking the degeneracy between the main lens and the perturber.}

Given this additional information and prior knowledge of the mass profiles of the main lens and of the perturber, many possible applications for the equations in Section~\ref{sec:equations} can be found, for example in the existing SLACS database \citep{bib:Bolton} or the SHARP database, which is currently under development  \citep{bib:Lagattuta}. \comm{Inserting specific models for the main lens and the perturber supported by observational evidence enables us to efficiently estimate bounds on potential perturbers because the approach allows us to analytically investigate the range of degeneracies of the properties of the perturbation, like its relative position along the line of sight, which to date has \newpage required elaborate simulations, as investigated in \cite{bib:Despali}.}

\begin{acknowledgements}
I thank Ashish K. Meena, Sven Meyer, Dominique Sluse, Volker Springel, Nicolas Tessore, Rüdiger Vaas, Simona Vegetti, Gerd Wagner, Dandan Xu, and the anonymous referee for helpful comments and discussions. I gratefully acknowledge the support from the Deutsche Forschungsgemeinschaft (DFG) WA3547/1-1.
\end{acknowledgements}

\bibliographystyle{aa}
\bibliography{aa}

\newpage
\appendix
\section{Derivation of Equations~\eqref{eq:le1} and \eqref{eq:le2}}
\label{app:derivations_les}

Starting with the Fermat potential
\begin{align}
\color{black}\phi(\boldsymbol{y}, \boldsymbol{x}) &\color{black}= \dfrac12 \left(\boldsymbol{x} - \boldsymbol{y} \right) - \psi(\boldsymbol{x}) \\
&\color{black}= \dfrac12 \left(r^2 + r_y^2 - 2 r_y r \cos(\theta_y - \theta) \right) - \psi(r, \theta) \;,
\end{align}
the lensing equations amount to
\begin{align}
\color{black}r_y \cos(\theta_y - \theta) &\color{black}= r - \dfrac{\partial \psi}{\partial r} = r - \alpha_r\\
\color{black}r_y \sin(\theta_y - \theta) &\color{black}= - \dfrac{1}{r} \dfrac{\partial \psi}{\partial \theta} = - \dfrac{\alpha_\theta}{r} \;,
\end{align}
denoting the derivatives of the potential by the deflection angle $\boldsymbol{\alpha} = (\alpha_r, \alpha_\theta)$.
For small perturbations from axisymmetry, $\left| \theta_y - \theta \right| \ll 1$, such that the approximation
\begin{align}
\color{black}r_y &\color{black}= r - \dfrac{\partial \psi}{\partial r} \label{eq:lea1} \\
\color{black}\theta_y &\color{black}= \theta - \dfrac{1}{r_y r} \dfrac{\partial \psi}{\partial \theta} \label{eq:lea2}
\end{align}
is valid.

Expanding Eq.~\eqref{eq:Taylor_potential} to second order
\begin{align}
\psi(r,\theta) = a_0 + p_0(\theta) + (a_1 + p_1(\theta)) (r - r_\mathrm{E}) + (a_2 + p_2(\theta)) (r - r_\mathrm{E})^2
\end{align}
and determining the derivatives with respect to the polar coordinates at most linear in $r-r_\mathrm{E}$ yields
\begin{align}
\dfrac{\partial \psi(r,\theta)}{\partial r} &= a_1 + p_1(\theta) + 2 (a_2 + p_2(\theta)) (r - r_\mathrm{E}) \;, \label{eq:app:psi_r} \\
\dfrac{\partial \psi(r,\theta)}{\partial \theta} &= \dfrac{\partial p_0(\theta)}{\partial \theta} +  \dfrac{\partial p_1(\theta)}{\partial \theta} (r - r_\mathrm{E}) \;. \label{eq:app:psi_theta}
\end{align}
Inserting Eqs.~\eqref{eq:app:psi_r} and \eqref{eq:app:psi_theta} in Eqs.~\eqref{eq:lea1} and \eqref{eq:lea2} and approximating $r_y \approx r$ in the denominator of Eq.~\eqref{eq:lea2} results in Eqs.~\eqref{eq:le1} and \eqref{eq:le2}.

\section{Derivation of Equations~\eqref{eq:ce1} and \eqref{eq:ce2}}
\label{app:derivations_ces}
Having two points $i$ and $j$ with the same radial source coordinate, I equate their Eqs.~\eqref{eq:le1}
to obtain
\begin{align}
r_i - r_j &= p_1(\theta_i) - p_1(\theta_j) + 2 a_2 (r_i - r_j)\nonumber \\
&\phantom{=} + 2\left( p_2(\theta_i)(r_i - r_\mathrm{E}) - p_2(\theta_j)(r_j - r_\mathrm{E}) \right) \\
&= \dfrac{p_1(\theta_i) - p_1(\theta_j)}{1-2a_2} \;,
\end{align}
where the last line is derived using $p_2 \ll a_2$, so that the terms in the second line can be neglected.

Rewriting the coefficients of the Taylor expansion in terms of potential derivatives (see Eq.~\eqref{eq:coefficients}) and using
\begin{align}
\left. \triangle \psi_\mathrm{a}(r) \right|_{r = r_\mathrm{E}} = 2 a_2 + 1 = 2 \kappa_\mathrm{a}(r_\mathrm{E}) \quad \Rightarrow 1-2a_2 = 2(1-\kappa_\mathrm{a}(r_\mathrm{E})) \;,
\end{align}
yields
\begin{align}
r_i - r_j &= \dfrac{1}{2(1-\kappa_\mathrm{a}(r_\mathrm{E}))} \left( \left. \dfrac{\partial  \psi_\mathrm{p}(\theta_i)}{\partial r} \right|_{r = r_\mathrm{E}} - \left. \dfrac{\partial  \psi_\mathrm{p}(\theta_j)}{\partial r} \right|_{r = r_\mathrm{E}} \right).
\end{align}
Replacing the potential derivatives by the radial deflection angle component $\alpha_{\mathrm{p},r} \color{black}{=\partial_r \psi_\mathrm{p}}$, Eq.~\eqref{eq:ce1} is obtained.

Eq.~\eqref{eq:ce2} is analogously derived by first subtracting Eq.~\eqref{eq:le2} for two differing angles $\theta_i, \theta_j$
\begin{align}
\theta_i - \theta_j &= \dfrac{1}{r_i\color{black}{^2}} \left. \dfrac{\partial  p_0(\theta)}{\partial \theta} \right|_{\theta=\theta_i} - \dfrac{1}{r_j\color{black}{^2}}\left. \dfrac{\partial p_0(\theta)}{\partial \theta} \right|_{\theta=\theta_j} \\
&\phantom{=} + \dfrac{1}{r_i\color{black}{^2}} \left. \dfrac{\partial  p_1(\theta)}{\partial \theta} \right|_{\theta=\theta_i} (r_i - r_\mathrm{E})- \dfrac{1}{r_j\color{black}{^2}} \left. \dfrac{\partial  p_1(\theta)}{\partial \theta} \right|_{\theta=\theta_j} (r_j - r_\mathrm{E}) \nonumber \\
&= \dfrac{1}{r_i\color{black}{^2}} \left. \dfrac{\partial  \psi_\mathrm{p}(\theta)}{\partial \theta} \right|_{\theta=\theta_i} -  \dfrac{1}{r_j\color{black}{^2}} \left. \dfrac{\partial \psi_\mathrm{p}(\theta)}{\partial \theta} \right|_{\theta=\theta_j} \;,
\end{align}
neglecting the terms in the second line and rewriting the coefficients as potential derivatives, as before. Replacing the latter by the angular component of the perturbing deflection angle $\alpha_{\mathrm{p},\theta}\color{black}{=\partial_\theta \psi_\mathrm{p}}$, Eq.~\eqref{eq:ce2} is obtained.

\section{Derivation of Equation~\eqref{eq:ntlo1}}
\label{app:ntlo}

Referring to the two-image configuration and notation of Figure~\ref{fig:configurations} (d), let the relative distance between the two source points of $(r_{A1}, \theta_A)$ and $(r_{A2}, \theta_A)$ be $\delta r_y$. Without loss of generality, the coordinate system in the source plane is rotated such that $\delta \theta_y = 0$.  Subtracting Equation~\eqref{eq:le1} for the two points then yields
\begin{align}
\delta r_y = r_{A1} - r_{A2} - 2(a_2 + p_2(\theta_A))(r_{A1} - r_{A2}) \;,
\end{align}
and analogously for the two points defining the width of arc $B$. Hence, \begin{align}
r_{A1} - r_{A2} &= \dfrac{\delta r_y}{1-2(a_2 + p_2(\theta_A))} \\
r_{B2} - r_{B1} &= \dfrac{\delta r_y}{1-2(a_2 + p_2(\theta_B))} \;,
\end{align}
which, using $1-2(a_2 + p_2(\theta_I)) \equiv 2(1-\kappa_I)$, $I=A, B$, yields Equation~\eqref{eq:ntlo1}.

Due to $\delta \theta_y = 0$, Equation~\eqref{eq:ce2} for $\theta_i = \theta_j = \theta_A$ 
\begin{align}
0 = \alpha_{\mathrm{p}, \theta}(r_\mathrm{E},\theta_A) \left( \dfrac{1}{r_{A1}\color{black}{^2}} - \dfrac{1}{r_{A2}\color{black}{^2}} \right) = \alpha_{\mathrm{p}, \theta}(r_\mathrm{E},\theta_B) \left( \dfrac{1}{r_{B2}\color{black}{^2}} - \dfrac{1}{r_{B1}\color{black}{^2}} \right)
\end{align}
can be employed to see that the coordinate system is chosen such that the perturbation deflection in angular direction vanishes.

\section{Derivation of Equation~\eqref{eq:perturbed_cc}}
\label{app:perturbed_cc}

As derived in \cite{bib:SEF}, the critical curve is determined by $\det (A) = 0$ with the magnification matrix
\begin{equation}
A (\boldsymbol{x}) = \left( \begin{matrix} 1 - \psi_{11}(\boldsymbol{x} )& -\psi_{12}(\boldsymbol{x})  \\ -\psi_{12} (\boldsymbol{x}) & 1 - \psi_{22}(\boldsymbol{x}) \end{matrix}\right) \;,
\end{equation}
where the derivatives of the deflection potential $\psi$ with respect to cartesian coordinates are denoted by subscripts $i,j=1,2$.
Expressing the cartesian second-order derivatives in polar coordinates 
\begin{align}
\psi_{11} &= \psi_{rr} \cos^2(\theta) + \psi_\theta \dfrac{\sin(2\theta)}{r^2} - \psi_{r \theta} \dfrac{\sin(2\theta)}{r} + \psi_r \dfrac{\sin^2 (\theta)}{r}\nonumber  \\
&\phantom{=} + \psi_{\theta \theta} \dfrac{\sin^2(\theta)}{r^2}\;, \\
\psi_{12} &=\psi_ {rr}  \sin(\theta) \cos(\theta) - \psi_\theta \dfrac{\cos(2\theta)}{r^2} - \psi_r \dfrac{\sin(2\theta)}{2r} + \psi_{r \theta} \dfrac{\cos(2\theta)}{r} \nonumber \\
&\phantom{=} - \psi_{\theta \theta} \dfrac{\sin(2\theta)}{2 r^2} \;, \\\psi_{22} &= \psi_{rr} \sin^2(\theta) - \psi_\theta \dfrac{\sin(2\theta)}{r^2} + \psi_{r \theta} \dfrac{\sin(2\theta)}{r} + \psi_r \dfrac{\cos^2 (\theta)}{r}\nonumber  \\
&\phantom{=} + \psi_{\theta \theta} \dfrac{\cos^2(\theta)}{r^2} \;, \\
\end{align}
and inserting the deflection potential of Equation~\eqref{eq:Taylor_potential} in
\begin{align}
\det(A) &= A_{11} A_{22} - (A_{12})^2 \\
 &= 1 - (\psi_{11} + \psi_{22}) + \psi_{11}\psi_{22} - (\psi_{12})^2
\end{align}
 yields the equation for the critical curve as
 \begin{equation}
 1 - \left(\psi_{rr} + \dfrac{\psi_{r}}{r} + \dfrac{\psi_{\theta \theta}}{r^2} \right) + \psi_{rr} \left( \dfrac{\psi_r}{r} + \dfrac{\psi_{\theta \theta}}{r^2}\right) - \dfrac{1}{r^2} \left( \dfrac{\psi_\theta}{r} - \psi_{r \theta} \right)^2 = 0 \;, 
 \end{equation} 
which can be approximated by 
\begin{equation}
 1 - \left(\psi_{rr} + \dfrac{\psi_{r}}{r} + \dfrac{\psi_{\theta \theta}}{r^2} \right) + \psi_{rr} \left( \dfrac{\psi_r}{r} + \dfrac{\psi_{\theta \theta}}{r^2}\right) = 0
\label{eq:app:cc}
\end{equation}
neglecting terms of order $r^{-4}$.
Using and deriving Equations~\eqref{eq:app:psi_r} and \eqref{eq:app:psi_theta} further, I obtain
\begin{align}
\dfrac{\partial \psi}{\partial r} &= a_1 + p_1(\theta) + 2 a_2 (r- r_\mathrm{E}) \;, \\
\dfrac{\partial \psi}{\partial \theta} &= \dfrac{\partial p_0 (\theta)}{\partial \theta} + \dfrac{\partial p_1 (\theta)}{\partial \theta}  (r-r_\mathrm{E}) \;, \\
\dfrac{\partial^2 \psi}{\partial r^2} &= 2 a_2 \;, \\
\dfrac{\partial^2 \psi}{\partial \theta^2} &=\dfrac{\partial^2 p_0 (\theta)}{\partial \theta^2} \;,
\end{align}
taking into account only terms containing second derivatives and terms linear in $r$.

Inserting these expressions into Equation~\eqref{eq:app:cc}, the equation for the critical curve reads
\begin{equation}
(1-2 a_2) r^2 - (a_1 - 2 a_2 r_\mathrm{E} + p_1 (\theta) ) r - \dfrac{\partial^2 p_0 (\theta)}{\partial \theta^2} = 0 \;.
\end{equation}
Replacing $a_1 = r_\mathrm{E}$ and $a_2 = \kappa_\mathrm{a} - 1/2$, Equation~\eqref{eq:perturbed_cc} is obtained.

\section{Derivation of Equation~\eqref{eq:time_delay}}
\label{app:time_delay}

The time delay between the arrival times of two images $A$ at $\boldsymbol{x}_A = (r_A, \theta_A)$ and $B$ at  $\boldsymbol{x}_B = (r_B, \theta_B)$, coming from the same source at $\boldsymbol{y} = (r_y, \theta_y)$, is given by
\begin{align}
\Delta t_{AB} &= t_A - t_B \;, \\
 &= \dfrac{\Gamma_\mathrm{d}}{c} \left( \dfrac12 \left( (\boldsymbol{x}_A - \boldsymbol{y} )^2 -(\boldsymbol{x}_B - \boldsymbol{y})^2\right) - \left( \psi(\boldsymbol{x}_A) - \psi(\boldsymbol{x}_B) \right) \right)
 \label{eq:time_delay_derivation}
\end{align}
with $\Gamma_\mathrm{d} =  D_\mathrm{d} D_\mathrm{s} /(D_\mathrm{ds}) (1+z_\mathrm{d})$. If we insert the perturbed deflection potential $\psi(\boldsymbol{x}) = \psi_\mathrm{a}(r)+ \psi_\mathrm{p}(r,\theta)$ and the consequently perturbed source position $\boldsymbol{y} + \delta \boldsymbol{y}$, we obtain
\begin{align}
\Delta t_{AB} = \dfrac{\Gamma_\mathrm{d}}{c} \left( \Delta \phi_\mathrm{a} - \delta \boldsymbol{y} \left(\boldsymbol{x}_A - \boldsymbol{x}_B \right) - \left( \psi_\mathrm{p}(\boldsymbol{x}_A) - \psi_\mathrm{p}(\boldsymbol{x}_B) \right) \right) \;,
\end{align}
to linear order in $\delta \boldsymbol{y}$, in which 
\begin{align}
\Delta \phi_\mathrm{a} = \dfrac12 \left( \left( \boldsymbol{x}_A - \boldsymbol{y} \right)^2 - \left(  \boldsymbol{x}_B - \boldsymbol{y}  \right)^2 \right) - \left( \psi_\mathrm{a}(\boldsymbol{x}_A) - \psi_\mathrm{a}(\boldsymbol{x}_B) \right) 
\end{align}
is proportional to the time delay caused by the main lens, the second term is the time delay caused by the source shift due to the perturbation and the last term is the time delay caused by the deflection potential of the perturber. Replacing the last term by $\Delta \psi_\mathrm{p}$, Eq.~\eqref{eq:time_delay} is obtained.

\end{document}